\documentclass[12pt]{article}

\usepackage{amsmath}
\usepackage{amssymb}
\usepackage{latexsym}
\usepackage{epic,eepic}

\usepackage[english]{babel}

\newcommand{\bbC}{{\mathbb C}}

\newtheorem{theorem}{Theorem}

\begin{document}

\title{Hyperdeterminants as integrable discrete systems}

\author{
  S.P.~Tsarev\thanks{SPT acknowledges partial
financial support from a grant of Siberian Federal University and
the RFBR grant 09-01-00762-a.} \   and T.~Wolf}

\date{}

\maketitle
\begin{center}
Siberian Federal University,\\
 Svobodnyi avenue, 79, \\
 660041,  Krasnoyarsk, Russia \\[1ex]
and \\[1ex]
Department of Mathematics, Brock University \\
500 Glenridge Avenue, St.Catharines, \\
Ontario, Canada L2S 3A1\\
e-mails: \\
\texttt{sptsarev@mail.ru} \ \ \
\texttt{twolf@brocku.ca}\\
\end{center}

\begin{abstract}
We give the basic definitions and some theoretical results about
hyperdeterminants, introduced by A.~Cayley in 1845. We prove
integrability (understood as $4d$-consistency) of a nonlinear
difference equation defined by the $2 \times 2 \times
2$ - hyperdeterminant. This result gives rise to the following
hypothesis: the difference equations defined by hyperdeterminants of
any size are integrable.

We show that this hypothesis already fails in the case of the
$2\times 2\times 2\times 2$ - hyperdeterminant.
\end{abstract}

\maketitle

\section{Introduction}\label{sec:intro}

Discrete integrable equations have become a very vivid topic in the
last decade. A number of important results on the classification of
different classes of such equations, based on the notion of
consistency \cite{BS}, were obtained in \cite{ABS,ABS2,TsWo08} (cf.\
also references to earlier publications given there). As a rule,
discrete equations describe relations on the scalar field variables
$f_{i_1 \ldots i_n}\in {\mathbb C}$ associated with the points of a
lattice ${\mathbb Z}^n$ with vertices at integer points in the
$n$-dimensional space ${\mathbb R}^n=\{(x_1,\ldots, x_n)| x_s \in
{\mathbb R}\}$. If we take the elementary cubic cell $K_n =
\left\{ (i_1, \ldots, i_n)\right. | $ $\left. i_s \in \{0,1\}
\right\}$ of this lattice and the field variables $f_{i_1 \ldots
i_n}$ associated to its $2^n$ vertices, an $n$-dimensional discrete
system of the type considered here is given by an equation of the
form
\begin{equation}\label{eq0}
 Q_n({\mathbf{f}}) = 0.
\end{equation}
Hereafter we use the short notation ${\mathbf{f}}$ for the
set $(f_{{00\ldots 0}}, \ldots , f_{11 \ldots 1})$ of all these
$2^n$ variables. For the other elementary cubic cells of ${\mathbb
Z}^n$ the equation is the same, after shifting the indices of
${\mathbf{f}}$ suitably.

The equations mostly investigated so far \cite{ABS,ABS2,TsWo08} were
supposed to have the following properties:

\textbf{1) Quasilinearity.} Equation (\ref{eq0}) is affine linear
w.r.t.\ every  $f_{i_1i_2 \ldots i_n}$,  i.e. $Q$ has degree 1 in
any of its four variables.

\textbf{2) Symmetry.} Equation (\ref{eq0}) should be invariant
w.r.t.\ the symmetry group of elementary cubic cell $K_n$ or its
suitably chosen subgroup.

On the other hand a number of interesting discrete equations which
do not enjoy one or both of these properties has been found. In this
publication we investigate an important class of symmetric discrete
equations which do not have the quasilinearity property and are
given by the equations $H_n(f_{{00\ldots 0}}, \ldots , f_{11 \ldots
1})=0$, were $H_n$ denotes the $n$-dimensional hyperdeterminant of
the corresponding $n$-index array $(f_{{00\ldots 0}}, \ldots , f_{11
\ldots 1})$. We give the precise definition of hyperdeterminants in
Section~\ref{sec:def}. In the simplest two-dimensional case of the
$2\times 2$ matrix the hyperdeterminant is just the usual
determinant:
\begin{equation}\label{2ddet}
    H_2({\mathbf{f}})= f_{00}f_{11}-f_{01}f_{10}.
\end{equation}
The next nontrivial case is the 3-dimensional $2\times 2\times
2$ - hyperdeterminant:
\begin{equation}\label{3ddet}
\begin{array}{rl}
  H_3({\mathbf{f}})=&
f_{111}^2 f_{000}^2 +f_{100}^2 f_{011}^2 +
f_{101}^2 f_{010}^2 +f_{110}^2 f_{001}^2 \\[0.3em]
  &-2 f_{111} f_{110} f_{001} f_{000}
      -2 f_{111}f_{101} f_{010} f_{000} \\[0.3em]
    &   -2 f_{111} f_{100} f_{011} f_{000} -2
f_{110} f_{101} f_{010} f_{001} \\[0.3em]
 & -2 f_{110} f_{100} f_{011} f_{001}
-2 f_{101} f_{100} f_{011} f_{010}\\[0.3em]
      &+4 f_{111} f_{100} f_{010}
f_{001} +4 f_{110} f_{101} f_{011} f_{000}.
\end{array}
\end{equation}

\begin{center}
\begin{figure}[http]
\setlength{\unitlength}{0.025em}
   \begin{minipage}[t]{170pt}
\begin{center}
\begin{picture}(220,220)(-80,0)
 \put(0,0){\circle*{20}}    \put(150,0){\circle*{20}}
 \put(0,150){\circle*{20}}  \put(150,150){\circle*{20}}
 \path(0,0)(150,0)       \path(0,0)(0,150)
 \path(150,0)(150,150)   \path(0,150)(150,150)
 \put(-40,-30){$f_{00}$}
 \put(-40,170){$f_{01}$}
 \put(155,-30){$f_{10}$}
 \put(155,170){$f_{11}$}
\end{picture}
\end{center}
\caption{Square $K_2$.}\label{fig2}
    \end{minipage}
\begin{minipage}[t]{200pt}
\begin{center}
\begin{picture}(200,220)(0,0)
 \put(0,0){\circle*{20}}    \put(150,0){\circle*{20}}
 \put(0,150){\circle*{20}}  \put(150,150){\circle*{20}}
 \put(50,50){\circle*{20}} \put(50,200){\circle*{20}}
 \put(200,50){\circle*{20}}
 \put(200,200){\circle*{20}}
 \path(0,0)(150,0)       \path(0,0)(0,150)
 \path(150,0)(150,150)   \path(0,150)(150,150)
 \path(0,150)(50,200)    \path(150,150)(200,200)
 \path(50,200)(200,200)
 \path(200,200)(200,50) \path(200,50)(150,0)
 \dashline[+30]{10}(0,0)(50,50)
 \dashline[+30]{10}(50,50)(50,200)
 \dashline[+30]{10}(50,50)(200,50)
 \put(-50,-30){$f_{000}$}
 \put(-50,170){$f_{001}$}
 \put(23,75){$f_{010}$}
 \put(22,220){$f_{011}$}
 \put(160,-25){$f_{100}$}
 \put(160,140){$f_{101}$}
 \put(215,50){$f_{110}$}
 \put(215,220){$f_{111}$}
\end{picture}
\caption{Cube $K_3$.}\label{fig3}
\end{center}
   \end{minipage}
\end{figure}
\end{center}

The corresponding elementary cells $K_2$, $K_3$ and the field
variables associated with the vertices are shown on
Figures~\ref{fig2},~\ref{fig3}.

The general definition of hyperdeterminants was given by
A.~Cayley~\cite{Cayley1}, who also gave the explicit
form~(\ref{3ddet}) of the first nontrivial $2\times 2\times
2$ - hyperdeterminant. In the last decades, following the modern and
much more general approach of $\mathcal{A}$-discriminants~\cite{GKZ},
the theory of hyperdeterminants found important applications in
quantum informatics \cite{MW02}, biomathematics \cite{AA00}, numerical
analysis and data analysis \cite{dSL06} as well as other fields.

As one can easily see, the expressions (3.7) in \cite{Kash} and
(6.11) in \cite{Schief}, describing some discrete integrable
equations, are nothing but the classical Cayley's $2\times 2\times
2$ - hyperdeterminant~(\ref{3ddet}). We prove below in
Section~\ref{sec:pmap} that (\ref{3ddet}) is also integrable in the
sense of \textit{$(n+1)$-dimensional consistency} \cite{BS}:

\textit{An $n$-dimensional discrete equation (\ref{eq0}) is called
consistent, if it may be imposed in a consistent way on all
$n$-dimensional faces of a $(n+1)$-dimensional cube}.

We give the accurate formulation of this general consistency
principle for the case of non-quasilinear expressions similar to
(\ref{3ddet}) in Section~\ref{sec:pmap}. For the two-dimensional
determinant (\ref{2ddet}) (which is quasilinear) consistency can be
established by a trivial computation; the equation
$H_2(f_{00},f_{11},f_{01},f_{10})=0$ is obviously linearized by the
substitution $f_{ij}=\exp \tilde f_{ij}$. Using a result on
Principal Minor Assignment Problem proved in \cite{HoSt} we
establish $4d$-consistency of the $2\times 2\times
2$ - hyperdeterminant (\ref{3ddet}) in Section~\ref{sec:pmap}, cf.\
Theorem~\ref{34dcons} below for the precise formulation.

This result gives rise to the following \emph{Conjecture}: the
difference equations defined by hyperdeterminants of any size are
integrable in the sense of $(n+1)$-dimensional consistency.
Nevertheless as we show in Section~\ref{sec:2222}, this Conjecture
fails already in the case of the $2 \times 2 \times 2 \times
2$ - hyperdeterminant. The computation of this $4d$ - hyperdeterminant
turns out to be highly nontrivial (compared to the relatively simple
expressions (\ref{2ddet}), (\ref{3ddet})) and was completed only
recently \cite{Sturm4}. We report in Section~\ref{sec:2222} a more
straightforward and simpler computation of the same hyperdeterminant
with the free symbolic computation program {\sc Form} \cite{FORM}. The
size of this hyperdeterminant ($2\,894\,276$ terms, total degree 24,
degree 9 w.r.t.\ each of the field variables) implies that checking
its $5d$-consistency can be done only numerically, using high
precision computation of roots of respective polynomial equations on
the $4d$-faces of the 5-dimensional cube $K_5$. This was done using
two different computer algebra systems {\sc Reduce} \cite{REDUCE} and
{\sc Singular} \cite{Sing}. As our computations have shown (cf.\ their
description in Section~\ref{sec:2222}), the $4d$ - hyperdeterminantal
equation $H_4(\mathbf{f})=0$ is not $5d$-consistent. This
non-integrability result should be investigated further since recent
examples \cite{LY} show that consistency is not the only possible
definition for discrete integrability.

\section{The definition of  hyperdeterminants and its
   variations}\label{sec:def}

The remarkable definition of hyperdeterminants given by A.~Cayley in
1845 \cite{Cayley1} and still used today \cite{GKZ} describes the
condition of singularity of an appropriate hypersurface. Let
$A=(a_{i_1i_2\cdots i_r})$ be a hypermatrix (an array with $r$
indices) with $i_s=0, \ldots , n_s$. The polylinear form
$$U=\sum_{i_1\cdots i_r}a_{i_1\cdots i_r}x^{(1)}_{i_1}
\cdots x^{(r)}_{i_r}$$ defines a hypersurface $U=0$ in $\bbC
P^{n_1}\times \ldots \times \bbC P^{n_r}$. Here $x^{(k)}_{i_k}$
denote the homogeneous coordinates in the respective complex
projective space $\bbC P^{n_k}$. This hypersurface is singular,
i.e.\ has at least one point where the condition of smoothness is
not satisfied iff the following set of $(n_1+1)\cdot \ldots \cdot
(n_r+1)$ equations
\begin{equation}\label{h-def}
    \left\{ \forall s=1,\ldots ,r, \quad \forall k=1, \ldots , n_s,\quad
\frac{\partial U}{\partial x^{(k)}_{i_s}}=0\right\}
\end{equation}
has a nontrivial solution $x^{(k)}_{i_s}\in \bbC P^{n_k}$. As one
can show (cf.\ \cite{GKZ}), if a certain condition (\ref{h-cond}) on
the dimensions $n_k$ of the array $A$ is satisfied, elimination of
the variables $x^{(k)}_{i_s}$ from (\ref{h-def}) results in a single
polynomial equation in the array elements $a_{i_1i_2\cdots i_r}$:
$H_r(A)=0$. This polynomial is irreducible and enjoys practically
the same symmetry properties as the usual determinant of a square
matrix. Following Cayley this polynomial $H_r(A)$ (defined uniquely
up to a constant factor) is called the \emph{hyperdeterminant of the
array $A$}. The necessary and sufficient condition of existence of a
\emph{single} polynomial condition $H_r(A)=0$ for the hypersurface
$U=0$ to be singular, i.e.\ the condition for the corresponding
hyperdeterminant of $A$ to be correctly defined, is as follows:
\begin{equation}\label{h-cond}
     \forall k, \qquad n_k \leq \sum_{s\neq k}n_s.
\end{equation}
In particular, if $r=2$, so for usual $(n_1+1)\times
(n_2+1)$-matrices, this condition implies $n_1=n_2$, and in this
case the hyperdeterminant $H_2$ coincides with the classical
determinant of the matrix $A_{i_1i_2}$. Note that for a given set
$\{n_1,\ldots,n_r\}$ of array dimensions one says that we have the
corresponding $(n_1+1)\times \ldots \times (n_r+1)$ - hyperdeterminant
since the array indices range from 0 to $n_k$.
 The hyperdeterminant is $SL(\bbC,n_1+1)\times \cdots \times
SL(\bbC,n_r+1)$-invariant, which means that if one adds to one
\emph{slice} $A_{k,p}=\{(a_{i_1i_2\cdots i_r}) \ \ \mathrm{with\
fixed}\ \ i_k=p\}$ another parallel slice $A_{k,q}$, $q \neq p$,
multiplied by some constant $\lambda$, the value of $H_r$ is
unchanged; swapping the slices $A_{k,p}$, $A_{k,q}$ either leave
$H_r$ again invariant or changes its sign depending on the parity of
the dimensions $n_i$; finally, multiplication of a slice $A_{k,p}$
with a constant $\lambda$ results in multiplication of the
hyperdeterminant by an appropriate power of $\lambda$. $H_r$ is also
invariant w.r.t.\ the transposition of any two indices $i_l$, $i_m$
of the hypermatrix $A=(a_{i_1i_2\cdots i_r})$.

As we have stated in the introduction, the first nontrivial $2\times
2\times 2$ - hyperdeterminant (\ref{3ddet}) was computed by
A.~Cayley himself \cite{Cayley1}. Amazingly enough, already the next
step, computation of the $2\times 2\times 2\times 2$ -
hyperdeterminant is very difficult. The problem of computation of an
\emph{explicit} polynomial expression for this case was proposed by
I.~M.~Gel'fand in his Fall 2005 research seminar at Rutgers
University. The monomial expansion of the $2\times 2\times 2\times
2$ - hyperdeterminant is related to some combinatorial problems, and
was done (using an inductive algorithm of L.~Schl\"afli~\cite{Schl})
for the first time in \cite{Sturm4}, using a dedicated C code; this
computation required a serious programming effort since the standard
computer algebra systems like Maple can not cope with the
intermediate large expressions. The resulting polynomial expression
for the $2\times 2\times 2\times 2$ - hyperdeterminant has
$2\,894\,276$ terms, total degree 24, and has degree 9 w.r.t.\ each
of the array entries $a_{i_1i_2i_3i_4}$. The size of this expression
in usual text format is around 200 megabytes.

In October 2007 we re-checked this computation of the $2\times 2\times
2\times 2$ - hyperdeterminant using a free symbolic computation program
{\sc Form} \cite{FORM} and the same inductive algorithm of
L.~Schl\"afli~\cite{Schl}. The computation required 8 hours on a 3GHz
processor of one of the SHARCNET nodes ({\tt www.sharcnet.ca}) and
some 800~Mb of temporary disk storage. Due to the efficient design of
{\sc Form} one had no need to write any special low-level code, the standard
{\sc Form} routines are completely sufficient. The obtained expression can
be saved either in text format (around 200~Mb) or in the internal
binary {\sc Form} format; both can be used as input to other {\sc Form} runs and
require only around 1 minute to be read into a session. Moreover, {\sc Form}
could cope with a straightforward check of the invariance of the
obtained hyperdeterminant w.r.t.\ addition of one slice $A_{k,p}$
multiplied by a symbolic constant $\lambda$ to another parallel slice
$A_{k,q}$, $q \neq p$.

This more
challenging computation still used the usual {\sc Form} routines and
required some 10 hours of CPU time and around 200~Gb of temporary
disk storage (the size of the intermediate expression reached $\sim
800\,000\,000$ terms!). This may suggest that some other
hyperdeterminants of higher size may also be computed using
available software and hardware.

Amazingly enough, if one reads the original Cayley papers
\cite{Cayley1}, then a \emph{relatively small} expression with some
340 terms which Cayley describes as \emph{the} $2\times 2\times
2\times 2$ - hyperdeterminant can be found! A straightforward check
shows that this expression enjoys the invariance properties for
hyperdeterminants stated above. On the other hand, Proposition~1.6
in \cite[p.~447]{GKZ} states that if a polynomial in the entries
$a_{i_1i_2\cdots i_r}$ of a hypermatrix $A$ has these invariance
properties \emph{and meets some extra weak condition on the stars of
the monomial powers} then it should be divisible by the respective
hyperdeterminant of $A$. The Cayley's expression does satisfy the
necessary invariance conditions; only a few terms do not meet the
required extra condition on the stars of the monomial powers, so
this statement of Cayley on the explicit form of the $2\times
2\times 2\times 2$ - hyperdeterminant is wrong. This was already
remarked by L.~Schl\"afli~\cite{Schl} who gave an inductive
algorithm for the computation of hyperdeterminants. Unfortunately, as
shown in \cite{WZ}, the Schl\"afli algorithm works only for very
special hyperdeterminants, in particular the only hyperdeterminants
with $n_1=\ldots=n_r=1$ which it can compute are precisely the
$2\times 2\times 2$- and $2\times 2\times 2\times
2$ - hyperdeterminants. Already for the $2\times 2\times 2\times
2\times 2$ - hyperdeterminant there seems to be no better way to
compute the explicit expression other than the elimination procedure
given in the definition of hyperdeterminants.

Based on the above symmetry discussion one could adopt other
definitions for hyperdeterminants.
Historically this resulted in a few other definitions of
hyperdeterminants as invariant expressions. Many of them have a much
simpler form than the definition we apply. A review of various
definitions can be found in \cite{Sokol}.

\section{The Cayley $2\times 2\times 2$ - hyperdeterminant as a discrete
integrable system and the Principal Minor Assignment Problem}\label{sec:pmap}

The integrability definition applied in this publication is based on
the requirement of consistency which is described in this section in
detail for the case of the $3d$ - hyperdeterminant (\ref{3ddet}).
Suppose we have a $4d$-cube $K_4$ shown on figure~\ref{fig4} with
field values $f_{ijkl}$, $i,j,k,l \in \{0,1\}$. One should impose
the formula (\ref{3ddet}) on every $3d$-face of $K_4$, by fixing one
of the indices $i,j,k,l$, and making it 0 for the faces which we
will call below ``initial faces'', or respectively 1 for the faces
which we will call ``final faces''. Further one needs to fix some
mapping from the initial ``standard'' $3d$-cube shown on
figure~\ref{fig3a} (with the vertices labelled $f_{ijk}$) onto every
one of the eight $3d$-faces (for example $\{f_{i1kl}\}$ on
$\{x_2=1\}$). Due to the symmetry properties of the $3d$ -
hyperdeterminant (\ref{3ddet}) this can be done, for example, using
the trivial lexicographic correspondence of the type $f_{ijk}
\mapsto f_{i1jk}$. The initial data are some arbitrary complex
values of the field variables $f_{ijkl}$ assigned to the vertices
shown on figure~\ref{fig4} by black circles. Then, using the
equation (\ref{3ddet}) imposed on the initial faces we can find the
values of the field variables for the last ($8^{\rm th}$) vertex of
the respective initial face, such vertices are shown on
figures~\ref{fig3a},~\ref{fig4} as white circles. Obviously since
the equations (\ref{3ddet}) for these values are quadratic we obtain
two possible values. On the next step we impose (\ref{3ddet}) to
hold on the 4 final faces using the initial data and the values
found on the previous step. From each of the 4 final $3d$-faces we
again obtain possible values for the final vertex $f_{1111}$ shown
on figure~\ref{fig4} by a small white box. For generic initial data
we have on \emph{each} final face $2^3=8$ different choices for the
intermediate values (white circles) so we find in principle $8\cdot
2 =16$ different possible values for $f_{1111}$ from each final
face. How many of them coincide among the 4 final faces?

\begin{center}
\begin{figure}[htbp]
\setlength{\unitlength}{0.025em}
\begin{minipage}[t]{200pt}
\begin{picture}(200,150)(-220,-20)
 \path(0,0)(150,0)       \path(0,0)(0,150)
 \path(150,0)(150,150)   \path(0,150)(150,150)
 \path(0,150)(50,200)    \path(150,150)(200,200)
 \path(50,200)(200,200)
 \path(200,200)(200,50) \path(200,50)(150,0)
 \dashline[+30]{5}(0,0)(50,50)
 \dashline[+30]{5}(50,50)(50,200)
 \dashline[+30]{5}(50,50)(200,50)
 \put(0,0){\circle*{12}}    \put(150,0){\circle*{12}}
 \put(0,150){\circle*{12}}  \put(150,150){\circle*{12}}
 \put(50,50){\circle*{12}} \put(50,200){\circle*{12}}
 \put(200,50){\circle*{12}}
 \put(200,200){\circle{12}}
 \put(-50,-30){$f_{000}$}
 \put(-50,170){$f_{001}$}
 \put(23,75){$f_{010}$}
 \put(22,220){$f_{011}$}
 \put(160,-25){$f_{100}$}
 \put(160,140){$f_{101}$}
 \put(215,50){$f_{110}$}
 \put(215,220){$f_{111}$}
\end{picture}
\caption{Cube $K_3$.}\label{fig3a}
\end{minipage}
\ \
   \begin{minipage}[t]{150pt}
\setlength{\unitlength}{0.045em}
\begin{picture}(200,150)(-70,-20)
 \put(  0,  0  ){\circle*{6}} \put(39.5, 15 ){\circle*{6}}
 \put(108, -5  ){\circle*{6}} \put(147 , 10 ){\circle{6}}

 \put(  0,114  ){\circle*{6}} \put(39.5,129 ){\circle{6}}
 \put(108,109  ){\circle{6}}
                              \put(144 ,121 ){\framebox(6,6){}}
 \put(49.5,41.5){\circle*{6}} \put(85  ,40  ){\circle*{6}}
 \put(62.5,46.5){\circle*{6}} \put(98  ,45  ){\circle*{6}}

 \put(49.5,80  ){\circle*{6}} \put(85  ,78.5){\circle*{6}}
 \put(62.5,85  ){\circle*{6}} \put(98  ,83.5){\circle{6}}

 \dashline[+15]{2.5}(  0,  0  )( 39.5, 15)
 \dashline[+15]{2.5}( 39.5, 15)(147, 10  )
 \path(108, -5  )(147, 10  ) \path(108, -5  )(  0,  0  )

 \path(  0,114  )(39.5,129 ) \path(39.5,129 )(147,124  )
 \path(147,124  )(108,109  ) \path(108,109  )(  0,114  )

 \path(  0,  0  )(  0,114  ) \dashline[+15]{2.5}(39.5, 15 )(39.5,129 )
 \path(108, -5  )(108,109  ) \path(147 , 10 )(147 ,124 )

 \path(49.5,41.5)(85  ,40  ) \path(85  ,40  )(98  ,45  )
 \dashline[+30]{2.5}(98  ,45  )(62.5,46.5)
 \dashline[+60]{2.5}(49.5,41.5)(62.5,46.5)

 \path(49.5,80  )(85  ,78.5) \path(85  ,78.5)(98  ,83.5)
 \path(98  ,83.5)(62.5,85  ) \path(62.5,85  )(49.5,80  )

 \path(49.5,41.5)(49.5,80  ) \path(85  ,40  )(85  ,78.5)
 \path(98  ,45  )(98  ,83.5) \dashline[+30]{2.5}(62.5,46.5)(62.5,85  )

 \path(  0,  0  )(49.5,41.5) \path(39.5, 15 )(62.5,46.5)
 \path(108, -5  )(85  ,40  ) \path(147 , 10 )(98  ,45  )
 \path(  0,114  )(49.5,80  ) \path(39.5,129 )(62.5,85  )
 \path(108,109  )(85  ,78.5) \path(147 ,124 )(98  ,83.5)

 \put(-20,-20){$f_{0001}$}  \put( 88,-25){$f_{1001}$}
 \put(-20,124){$f_{0011}$}  \put(155,10){$f_{1101}$}
 \put(5,37){$f_{0000}$}   \put(135,134){$f_{1111}$}
\end{picture}
\caption{Cube $K_4$.}\label{fig4}
    \end{minipage}
\end{figure}
\end{center}

As we prove below, 8 of every 16 values for $f_{1111}$ found for
each of the final $3d$-faces are common. All other $8\cdot 4=32$
are in general not shared between the final faces as it has been confirmed
by numerical examples.

This result should be considered as the proof of consistency for
face formula (\ref{3ddet}); our considerations below are based on
a remarkable result proved in \cite{HoSt}.

Let us first formulate the necessary definitions and results of
\cite{HoSt}. Suppose we have a real symmetric $n{\times}n$-matrix
$M=(m_{ij})$. Its principal minors form a vector of length $2^n$
with entries indexed by subsets $I$ of the set $\{1,2,\ldots,n\}$.
Namely, $M_{I}$ denotes the minor of $M$ whose rows and columns are
indexed by $I$. This includes the $0 {\times} 0$-minor
$M_{\emptyset} = 1$. The famous Principal Minor Assignment Problem
considers the description of a suitable complete set of algebraic
relations among the minors of a generic symmetric
$n{\times}n$-matrix $M=(m_{ij})$. The first observation (formula
(2)) of \cite{HoSt} consists in the fact that for principal minors
of a $3\times 3$ symmetric matrix one has the following relation:
$$     \begin{array}{l}
   M_{\emptyset}^2 M_{123}^2  +
 M_{1}^2 M_{23}^2 +   M_{2}^2 M_{13}^2 + M_{3}^2 M_{12}^2
 + 4  M_{\emptyset} M_{12} M_{13} M_{23} +
 4  M_{1} M_{2} M_{3} M_{123}
\\    -  2   M_{\emptyset}   M_{1} M_{23} M_{123}
 -  2   M_{\emptyset}   M_{2} M_{13} M_{123}
 -  2  M_{\emptyset}   M_{3} M_{12} M_{123}
 -  2  M_{1} M_{2} M_{13} M_{23} \\
 -  2  M_{1} M_{3} M_{12} M_{23}
 - 2   M_{2} M_{3} M_{12} M_{13}
 \quad = \quad 0 .  \end{array}
$$
This obviously gives us the Cayley's hyperdeterminant (\ref{3ddet})
if we interpret every minor $M_I$ as the field variable
$f_{i_1i_2i_3}$ with $i_s=1$ if $s \in I$ and  $i_s=0$ otherwise,
for example $M_{13}=f_{101}$. For the  initial vertex we have
$f_{000}=M_{\emptyset} = 1$. For symmetric matrices $M$ of larger
size the $2 \times 2 \times 2$ - hyperdeterminantal relations are also
fulfilled for ``shifted'' principal minors, in our terminology this
means that the hyperdeterminant (\ref{3ddet}) vanishes on every
$3d$-face of the $n$-dimensional hypercube with field variables
$f_{i_1 \ldots i_n}$ equal to the principal minors $M_I$ such that
$i_s=1$ if $s \in I$ and $i_s=0$ otherwise. As a remarkable fact
(not necessary to us) we mention that for $d>2$ on any
$d$-dimensional face of this $n$-dimensional hypercube the corresponding
$d$-dimensional hyperdeterminant also vanishes:

\begin{theorem}[\cite{HoSt}] Let $M = (m_{ij})$ be a symmetric
$n{\times}n$ matrix.  Then the vector $M_*$ of all principal minors of
$M$ is a common zero of all the hyperdeterminants of formats from $2
{\times} 2 {\times} 2$ up to
${\underbrace{2{\times}2{\times}\cdots{\times}2}_{n \; terms}}$.
\end{theorem}

For $n \geq 2$,  the entries of the symmetric matrix $M = (m_{ij})$
are determined up to sign by their principal minors of size
$1{\times}1$ and $2{\times}2$ since $m_{ii} = M_{i}$, $m_{ij}^2 =
M_{i} M_{j} - M_{ij} M_\emptyset$. As we see, the value
$f_{111}=M_{123}$ for the final vertex is not defined uniquely from
the relation (\ref{3ddet}); the choice of one of the two possible
values corresponds to a suitable choice of the signs for the 3
off-diagonal elements $m_{12}$, $m_{13}$, $m_{23}$. These $2^3=8$
sign combinations give only two different values for $M_{123}$,
because changing simultaneously the signs of a certain row and the
column symmetric to it we do not change the principal minors of the
matrix, so we can fix the signs of $m_{12}$ and $m_{13}$.

If we take a $4 \times 4$ symmetric matrix, all its elements will be
defined up to the sign of the off-diagonal elements by the principal
minors of size $1{\times}1$ and $2{\times}2$; this corresponds
precisely to the choice of the initial data (black circles) on
figure~\ref{fig4}. So provided we choose the initial data $f_{ijkl}$
with $f_{0000}=1$, we can find a corresponding $4 \times 4$
symmetric matrix with fixed $1{\times}1$ and $2{\times}2$ minors;
the off-diagonal elements of the matrix are fixed up to the signs.
So, fixing 3 signs of the 6 off-diagonal elements (changing
simultaneously the signs of a certain row and the column symmetric
to it) we have only $3$ essential sign choices; as the numerical
examples show, they really give $2^3=8$ different minors
$M_{1234}=\det M$. Precisely these 8 values should give the 8 values
for $f_{1111}$ that should coincide after computation of 16 possible
candidates for $f_{1111}$ from each of the 4 final faces from the
initial data.

If we choose initial data with $f_{0000} \neq 1$, $f_{0000} \neq 0$,
homogeneity of the face equations (\ref{3ddet}) allows us to reduce
this situation to the case $f_{0000} = 1$ considered above. Thus the
following statement has been proved:

\begin{theorem}\label{34dcons} Let some generic initial data
$\big\{f_{0000},$ $f_{1000},$ $f_{0100},$ $f_{0010},$ $f_{0001},$
$f_{1100},$ $f_{1010},$ $f_{1001},$ $f_{0110},$ $f_{0101},$
$f_{0011}\big\}$ on the cube $K_4$ be given.  After computation of
the two possible values for each of the intermediate vertices
$\big\{f_{1110}, f_{1101}, f_{1011}, f_{0111}\big\}$ from the face
relations (\ref{3ddet}) on the respective $3$-dimensional initial
faces, among the sets of 16 possible values of $f_{1111}$ for each
if the 4 final faces, subsets of 8 values coincide for all of them.
They are equal to the 8 possible values of $\det M$ for the
symmetric matrices having given $1{\times}1$ and $2{\times}2$
principal minors corresponding to the initial data for $f_{ijkl}$.
\end{theorem}

\section{The next step: the $2\times 2\times 2\times 2$ - hyperdeterminant
and its $5d$-inconsistency}\label{sec:2222}

The definition of hypothetical $5d$-consistency for the $2\times
2\times 2\times 2$ - hyperdeterminant which has degree 9 w.r.t.\ each
of its 16 variables is easily formulated along the lines given in
the previous Section. So in this case not only the size of the face
equations, but also the number of possible choices of the
intermediate values $f_{ijklm}$ with $i+j+k+l+m=4$ for the
computation of the final $f_{11111}$ from each of the 5 final faces
of the 5-dimensional hypercube $K_5$ is dramatically increased.

The strategy of numerical checking the hypothesis of
$5d$-consistency adopted by us involved the following steps.

\smallskip

1) We assign some random integer values for the initial data
$f_{ijklm}$ with $i+j+k+l+m \leq 3$ and use {\sc Form} to substitute them
into the expressions of the $2\times 2\times 2\times
2$ - hyperdeterminant on the 5 initial $4d$-faces obtaining univariate
polynomials for each of the intermediate $f_{01111}$, \ldots ,
$f_{11110}$ and output the resulting expressions into a text file for
further processing by {\sc Reduce} and independently by {\sc Singular}. For
initial data being random integers in the range $[1,100]$, the
obtained equations have integer coefficients with approximately 40
decimal digits.

\smallskip

2) The same substitution of the initial data into the final faces is
performed with {\sc Form}, resulting in much larger multivariate
polynomials for the intermediate values $f_{01111}$, \ldots ,
$f_{11110}$ and the final $f_{11111}$. These polynomials (each has
the size of $\sim \! 360$~Kb) are output into a text file for
further processing by {\sc Reduce} and independently by {\sc Singular}.

\smallskip

3) The 5 univariate polynomial equations for the intermediate
$f_{01111}$, \ldots , $f_{11110}$ are solved with a guaranteed
precision of 150 digits.

\smallskip

4) For each one of the $9^5$ combinations of the 9 complex roots the
following computation is performed.

\smallskip

a) The set of complex roots for $f_{01111}$, \ldots , $f_{11110}$ is
replaced in the 5 final face relations obtained on step 2, which
makes them univariate polynomials for $f_{11111}$ with complex
rounded coefficients.

\smallskip

Starting with a guaranteed precision of 20 digits:

\smallskip

b) One of the obtained polynomials is solved for $f_{11111}$ and

\smallskip

c) successively the other 4 polynomials are solved for $f_{11111}$
as long as there is a non-empty {\em approximate intersection} (with
a definite relative tolerance, see below) of the sets of roots for
the $f_{11111}$ for all the polynomials solved so far.

\smallskip

d) If all five face relations have at least one common approximate
solution then execution continues with step b) with twice as many
guaranteed precise digits, up to a maximum of 80 digits. This was
never necessary.

To increase safety, two complex values $u, v$ were only
considered NOT to be approximately equal if for $p$ precise digits the
difference $u-v$ differed significantly from zero, more precisely if
$|u-v| / |u| > 10^{-p/2}$.

\bigskip

The computation performed on the nodes of SHARCNET has shown that
\emph{no equal values for $f_{11111}$} are obtained from the 5 final
faces. The details of the computation and the code used can be
obtained from the authors or downloaded from
\texttt{http://lie.math.brocku.ca/twolf/papers/TsWo2008/}.

\bigskip

This results in the conclusion that:

\emph{The $2\times 2\times 2\times 2$ - hyperdeterminant is not
$5d$-consistent}.

\bigskip

The safety of our numerical inconsistency result is increased by the
fact that it is obtained by two completely different computer
algebra systems. The {\sc Singular} package is written in C using an
arbitrary precision C library for their numerical computations
whereas {\sc Reduce} uses a long number arithmetic implemented in {\sc Lisp}.

We have also checked that the relatively small expression given by
Cayley \cite{Cayley1} for the $2\times 2\times 2\times
2$ - hyperdeterminant is also $5d$-inconsistent.

\section*{Acknowledgements}

The authors enjoy the opportunity to thank A.Bobenko, O.Holtz,
B.Sturmfels and Yu.Suris for fruitful discussions, references to
original sources and valuable remarks.

For this work facilities of the Shared Hierarchical Academic
Research Computing Network (SHARCNET: {\tt www.sharcnet.ca}) were
used. SPT acknowledges partial financial support via SHARCNET Senior
Visiting Fellowship Programme during his visit to Brock University in
October 2007 where the main part of this work has been completed.


\end{document}